# High-throughput quantification of interfacial protein load in monodisperse oil-in-water emulsions

Varun Potdar[1], Alexander Favela[1], Michael Boehm[2], Stefan Baier[3], and Siva Vanapalli[1]

[1]*Department of Chemical Engineering, Texas Tech University, Lubbock, TX, USA*

[2]*Motif FoodWorks,Inc., Boston, MA, USA*

[3]*School of Chemical Engineering, The University of Queensland, Brisbane, QLD 4072, Australia*

Email for correspondence: siva.vanapalli@ttu.edu

## Abstract

Food proteins play a critical role as emulsifiers, where adsorption at the oil–water interface governs emulsion formation and stability. However, quantitative characterization of interfacial protein load remains challenging due to polydispersity in conventional emulsification methods. Here, we present a microfluidic workflow combined with colorimetric protein quantification to study interfacial protein adsorption on monodisperse emulsions under uniform droplet sizes and environmental conditions. Interfacial load was measured immediately after emulsion collection to define the post-collection initial load ($\Gamma_o$) and subsequently monitored over 48 h to obtain the saturation load ($\Gamma_\infty$). Interfacial load exhibited a strong bulk concentration-dependent trend and was well described by a pseudo-first-order model. Both $\Gamma_o$ and $\Gamma_\infty$ followed sublinear scaling with concentration ($\Gamma_o \sim C^{0.21}$, $\Gamma_\infty \sim C^{0.12}$), indicating progressively smaller increases in interfacial load with increasing bulk protein concentration. Droplet-size analysis revealed largely diameter-independent interfacial loads, with modest early-loading reductions for the smallest droplets at low protein concentration. Comparison with idealized interfacial packing limits shows that measured loads exceed monolayer predictions and lie between bilayer and trilayer bounds, suggesting protein accumulation beyond a simplified monolayer model. Additionally, high-throughput mapping was performed across three proteins, seven concentrations, three pH values, and two ionic strengths, revealing that α-Casein exhibited higher interfacial loads than β-Casein and Whey protein Isolate, with maximum adsorption near the isoelectric point and at elevated ionic strength. In summary, this study establishes a quantitative experimental framework for linking protein properties, formulation conditions, and interfacial behavior using monodisperse microfluidic emulsions, while requiring only milligram-scale protein quantities.

## 1. Introduction

Dairy proteins play an important role in the formation and stabilization of various emulsion-based products in the food industry.[1-3] Their amphiphilic nature allows them to adsorb at interfaces and enhance emulsion stability.[4] Upon adsorption, proteins can reduce interfacial tension, alter the charge at the interface depending on pH and ionic strength and unlike traditional surfactants impart local viscoelastic properties.[5-9] The modifications at the interface depend significantly on the protein type and structure.[4, 10] Globular proteins, such as Whey Protein Isolate (WPI) and β-lactoglobulin (BLG), possess compact structures stabilized by strong intramolecular interactions, resulting in greater structural stability. This stability necessitates that these proteins undergo conformational changes to effectively adsorb at an interface, a process that can be relatively slow due to their compact structures, thereby contributing to slower adsorption rates.[11-13] In contrast, flexible proteins like α- and β-Casein, which lack rigid tertiary structures, can more readily adapt their conformations upon interacting with an interface, facilitating faster adsorption kinetics.[11-13] Understanding these behaviors is essential for optimizing emulsion stability and exploring specific protein characteristics.

The adsorption of proteins over time is a complex process, involving rapid initial adsorption, followed by gradual structural rearrangement and aggregation.[14, 15] Short-term adsorption studies, which typically last from milliseconds to hours, seek to understand interaction of individual proteins with the oil-water interface and ensuing structural conformations of the proteins as they absorb to the interfaces.[1, 16-25] Alternatively, long-term adsorption studies covering days or weeks explore the dynamic evolution of protein-protein interactions at interfaces, providing insights into slow processes like multi-layers, and their rearrangement.[14, 15, 26, 27] Characterizing these long-term adsorption dynamics is essential for understanding shelf-stability of emulsions and evaluating protein performance in product applications.[27] In both the short- and long-term adsorption studies, a key parameter of interest is the amount of protein adsorbed per unit area at the interface or the interfacial load ($\Gamma$). Established techniques, including interfacial tensiometry, interfacial rheology, quartz crystal microbalance (QCM), and light scattering, have been employed to investigate protein adsorption and interfacial behavior, providing complementary information on adsorption dynamics, interfacial mechanical properties, adsorbed mass, and protein organization[3, 28, 29]f[30]. In contrast, quantitative estimation of interfacial protein load in emulsified systems with well-defined droplet size and interfacial area remains experimentally challenging.

Protein adsorption at oil–water interfaces is strongly influenced by environmental conditions, including pH, ionic strength, and protein concentration, which collectively govern interfacial interactions and adsorption behavior. While intrinsic protein properties also play a role, as discussed above, electrostatic effects arising from pH and ionic strength are particularly important in determining adsorption. Proteins near their isoelectric point (pI) exhibit minimal net charge, reducing electrostatic repulsion and promoting adsorption, whereas at pH values far from the pI, increased charge leads to stronger repulsion and lower adsorption. Ionic strength further modulates these interactions by screening charges on protein molecules and at the interface, with higher ionic strength generally enhancing adsorption [31, 32]. However, systematic investigation of these variables using conventional emulsification techniques remains challenging. Traditional methods, such as high-pressure homogenization, typically require large sample volumes and involve simultaneous droplet formation and recoalescence, resulting in polydisperse emulsions that complicate quantitative analysis of adsorption dynamics and stability [33, 34].

A key consequence of droplet polydispersity is the difficulty in accurately estimating the total interfacial area available for protein adsorption [4, 35]. In conventional emulsions, interfacial load is typically inferred from mass balance approaches using an average droplet size, most commonly the Sauter mean diameter ($d_{32}$) [3, 4]. While this parameter provides a surface-area-weighted mean, it does not fully represent the breadth of the droplet size distribution present in the system. As a result, emulsions with similar reported mean diameters may still differ substantially in their total interfacial area, leading to variability in the calculated interfacial load[3, 4]. This variability can propagate into interpretation of adsorption behavior, making it difficult to distinguish whether observed differences arise from changes in protein properties or from differences in the underlying droplet population. This effect is particularly relevant in depletion-based measurements, where adsorption is determined from bulk concentration changes that reflect the collective contribution of all droplets present in the system[1, 18]. Consequently, reported interfacial load values may reflect not only intrinsic adsorption behavior but also differences in droplet size distributions. These limitations highlight the need for experimental approaches that provide well-defined droplet sizes and interfacial areas, enabling more reliable and reproducible quantification of protein adsorption.

Microfluidics offers an attractive platform for generating monodisperse emulsions with minimal requirements on sample volumes to investigate emulsion formation and stability at processing-relevant length and timescales.[34, 36-40] It provides precise control over droplet size and formation conditions, allowing for comprehensive analysis of interfacial tension, coalescence and adsorption kinetics.[41, 42] For instance, interfacial tension effects occurring at the millisecond timescale during droplet formation have been investigated using microfluidics.[43-45] Coalescence investigations in microfluidic channels, where emulsion droplets collide and potentially coalesce if not sufficiently stabilized have been studied, providing insights on dynamics of droplet stability.[46, 47] Another study using fluorescently labeled surfactants in the continuous oil phase revealed non-uniform adsorption layers at the droplet interface, emphasizing the influence of adsorption kinetics and surface convection on surfactant distribution.[48] Thus, microfluidic technologies are well suited for studying emulsions and related interfacial phenomena.

Most previous microfluidic studies involving adsorption kinetics have focused on small-molecule surfactants[38, 39, 46, 47, 49], and very few on proteins, such as WPI[49-51], β-lactoglobulin[50], casein[52], and soy proteins[52]. These studies focused on understanding the relationship between emulsion stability and adsorption timescales, where adsorption timescale was determined by the residence time of droplets within microchannels. Emulsion stability was assessed based on the frequency of droplet coalescence, with measurement timescales ranging from milliseconds to seconds[49-51] or in few instances up to 15 days[52]. However, direct quantification of adsorbed protein load at the interface over both short (e.g., minutes) and long (e.g., days) timescales remains limited. This gap underscores the need for new approaches to quantify protein adsorption kinetics at the interface over a wide range of timescales.

In this study, we combined droplet microfluidics with standard colorimetric analyses[53] to measure the amount of protein interfacially absorbed at monodisperse oil-in-water emulsions. Using this approach, we characterize the adsorption kinetics of four types of proteins: flexible α-Casein and β-Casein, and globular proteins such as whey protein isolate (WPI) and β-lactoglobulin (BLG). This approach enabled detailed evaluation of protein adsorption kinetics, capturing interfacial load immediately after droplets are collected and during long-term storage. The effect of environmental variables such as pH and ionic strength, as well as protein-specific parameters (e.g. protein type, and protein concentration) were systematically investigated.

## 2. Materials and methods

### i. Protein sample preparation

In this study, the following four different proteins were investigated due to their differences in molecular characteristics and functionalities (Table 1)[54-60]. Whey Protein Isolate (WPI, Hilmar Ingredients, CA, USA) – a mixture of amphoteric globular proteins, β – Lactoglobulin (Sigma-Aldrich, St. Louis, USA) a major constituent of whey protein with a well-defined globular structure, and α – Casein (Sigma-Aldrich, St. Louis, USA), and β – Casein (Sigma-Aldrich, St. Louis, USA), flexible and amphiphilic proteins.

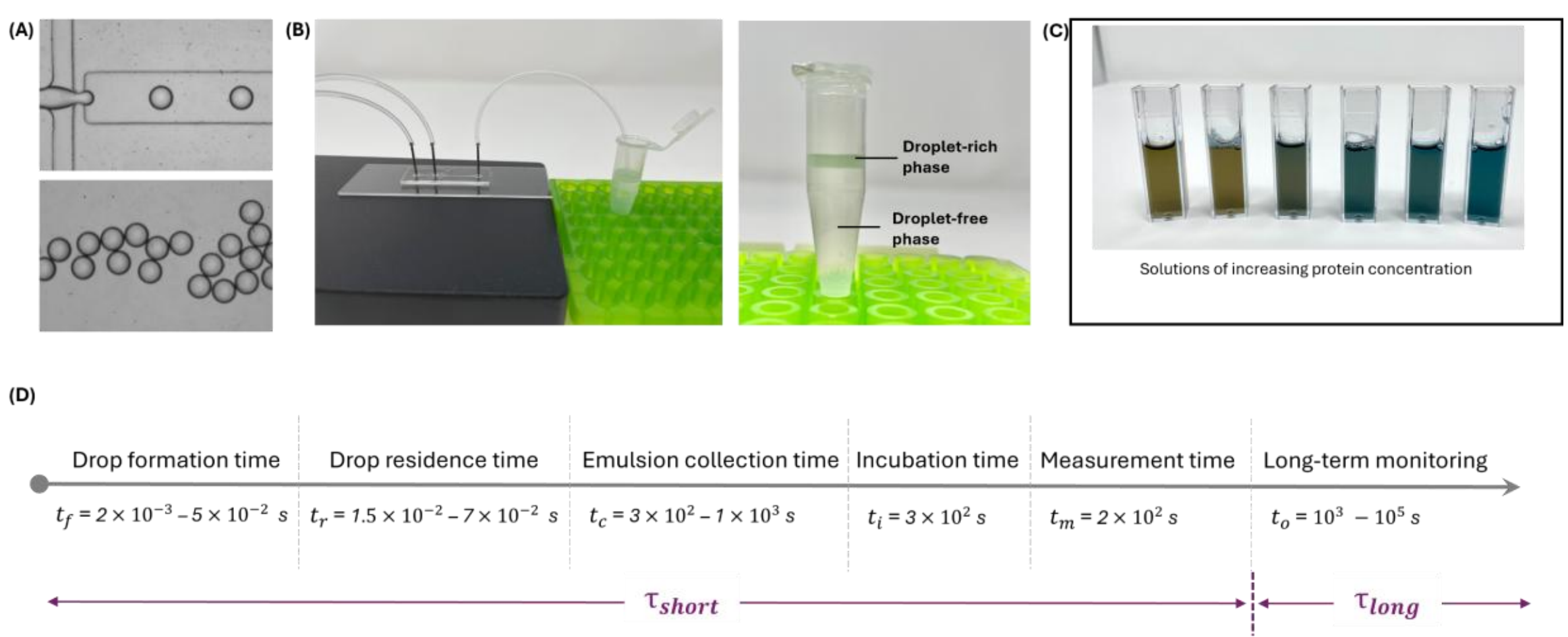

**Figure 1:** Experimental workflow for measuring adsorbed protein load and the associated timescales. A) Microfluidic flow focusing device geometry that produces monodisperse oil-in-water emulsions; B) Generated emulsions were collected with phase separates into a droplet-rich layer and a droplet-free continuous phase; C) Continuous phase collected after phase separation was mixed with the Bradford reagent and absorbance was measured at 595 nm. D)Timescales associated with each step of the experiment workflow. In this study, protein load at the interface was measured at timescale $\tau_{long}$ > 1000 s.

Buffer solutions were made with sodium phosphate monobasic (Sigma-Aldrich, St. Louis, USA) and sodium phosphate dibasic (Sigma-Aldrich, St. Louis, USA) to achieve three pH levels: 3, 5 and 7. The pH was adjusted with 0.05 M HCl and 0.05 M NaOH. The dispersed phase consisted of n-Hexadecane (Acros-Organics, New Jersey, USA). Protein solutions with varying concentrations were prepared in 5 mM phosphate buffer by mixing for 2 hours at room temperature (21 ± 1 °C). These solutions were then kept at 4°C and used within two days of preparation. Protein quantification was performed using Bradford reagent (Sigma-Aldrich, St. Louis, USA) as discussed further below.

| Proteins | Molecular weight (kDa) | Number of amino acids | Isoelectric point | Zeta Potential (mV) | Hydrodynamic radius (nm) |
|---|---|---|---|---|---|
| α-Casein | 23.5 | 214 | 4.6[51] | (-25 ± 1)[52] | 2.9[53] |
| β-Casein | 23.8 | 209 | 4.9[54] | (-34 ± 2)[54] | 3.7[53] |
| β-Lactoglobulin | 18.36 | 162 | 5.2[55] | (-31 ± 2)[56] | 3.5[57] |

**Table 1:** Physicochemical properties of α-Casein, β-Casein, and β-lactoglobulin, as reported in the literature [51-57].

### ii. Production of protein-stabilized emulsions

Protein-stabilized oil-in-water emulsions were generated using hydrophilic PDMS microfluidic flow focusing device, as shown in **Fig. 1A**. Fluid flow was controlled using syringe pumps. Droplet generation was operated in the squeezing regime, resulting in highly monodisperse emulsions under the chosen flow conditions. The dispersed phase flowrate ($Q_{disp}$) was fixed at 200 μL/hr, while the continuous phase flowrate ($Q_{cont}$) was set to 2000 μL/hr. The resulting oil volume fraction in the emulsions was approximately 9%, ensuring stable emulsions under given flow conditions. For the droplet-size experiments, the dispersed and continuous phase flow rates were varied to generate different droplet sizes while maintaining a constant oil volume fraction of approximately 9%.

Two techniques were employed for making protein-stabilized emulsions: (i) *no-dilution*, where droplets were formed at the same bulk protein concentration as the target (or intended) protein concentration, and (ii) *dilution*, where droplets were initially formed at the target bulk protein concentration of 0.1 wt% and then diluted to the respective lower protein concentrations. We found that the no-dilution approach led to wettability issues particularly at the lower protein concentration (e.g. below 0.025 wt%), making it difficult to produce monodisperse emulsions. In contrast, the dilution approach ensured the emulsion production was stable with uniform droplet size.

For experiments involving different droplet sizes, device geometries and flow conditions were adjusted to obtain the desired droplet diameters while maintaining monodispersity **(Fig. S1**). Various protein concentrations for each of the proteins were tested (**Table 2**). The formation of droplets at the orifice was observed using a high-speed camera (Phantom VEO-E 310L, Vision Research, NJ, USA) and an inverted microscope (ECLIPSE Ts2, Nikon Instrument Inc., NY, USA) at 10x magnification (resolution 1 pixel = 2 um).

| Protein | Concentration range (wt%) | Technique | Drop Diameter (µm) | Interfacial tension (mN/m) |
|---|---|---|---|---|
| α- Casein | 0.005 - 0.1 | Dilution | 78 - 85 | 12 - 43 |
| | | No dilution | | |
| β- Casein | 0.005 - 0.1 | Dilution | 78-85 | 16 - 45 |
| | | No dilution | | |
| β-Lactoglobulin | 0.5 - 3 | No dilution | 80-85 | 10 - 25 |
| Whey Protein Isolate | 0.01 - 0.1 | Dilution | 75 - 86 | 8 - 34 |
| | | No dilution | | |
| | 0.5 - 3 | No dilution | | |

**Table 2:** Experimental conditions of the proteins along with the measured average droplet size and interfacial tension.

The frequency and size of droplets were determined using high-speed video recordings. For each run, at least 1000 frames were recorded and processed with ImageJ software (version 1.53t) to provide precise measures of droplet size distribution and generation frequency. Following droplet generation, the emulsions were collected using an outlet tube attached to a microfluidic device that directed the emulsion into Eppendorf tubes **(Fig. 1B)**. A fixed volume of 1 mL was collected during a predetermined time period to ensure sampling uniformity.

### iii. Determination of adsorbed protein

The amount of protein adsorbed at the oil-water interface was quantified by considering the surface area of the monodisperse microfluidic oil droplets and the change in the protein concentration within the continuous phase after emulsification. The protein concentration in the continuous phase was evaluated using the Bradford method, which involved binding of Coomassie Brilliant Blue G250 dye to the protein, resulting in a detectable shift in the light absorption spectrum. First, the absorbance values of bulk protein solutions at different concentrations were measured using UV-vis spectrophotometry (UNICAM 5625) at a wavelength of 595 nm **(Fig. 1C)**, to generate linear calibration curves **(Fig. S2)**. Following emulsification, the protein absorbance values in the continuous phase were measured, and the amount of protein adsorbed, denoted as $\Gamma$ $(mg/m^2)$ at the oil-water interface was quantified using the formula.

$$\boldsymbol{\Gamma} = \frac{(C_{ini} - C_{ser})\, V_{aq}}{nA_{drop}} \qquad (1)$$

where $C_{ini}$ represents the initial protein concentration in the bulk solution (mg/mL). $C_{ser}$ represents the protein concentration in the continuous phase after emulsification (mg/mL). $V_{aq}$ is the volume of the aqueous phase in the collected emulsion, $A_{drop}$ is the surface area of a single drop, and n is the total number of droplets. The total number of droplets was calculated using the frequency of drop generation and the emulsion collection time.

As shown in **Fig. 1D**, protein adsorption can occur from the moment of droplet formation to droplets being collected, incubated and measured. This cumulative adsorption time is less than 30 minutes, which is much smaller than the 48-hr time-course over which interfacial protein load was measured. Specifically, absorbance readings were taken at time points of 1, 2, 4, 8, 12, 18, 24, 32, and 48 hours after the emulsion collection and incubation period.

## 3. Results

## i. Quantification of interfacial protein adsorption kinetics

We first assess the precision and reproducibility of interfacial protein load measurements enabled by the microfluidic platform. To quantify the kinetics of protein adsorption at the oil-water interface, the generated monodisperse droplets were collected and the interfacial load at the interface was quantified over time using the Bradford assay. Measurements were taken at regular intervals over a 48 hour period to capture the dynamics of protein adsorption. To begin with, the protein adsorption was characterized using bovine α-Casein across different bulk protein concentrations ranging from 0.005 wt% to 0.1wt%. Emulsions were produced using both the dilution and no-dilution approaches to address potential wettability effects during emulsification.

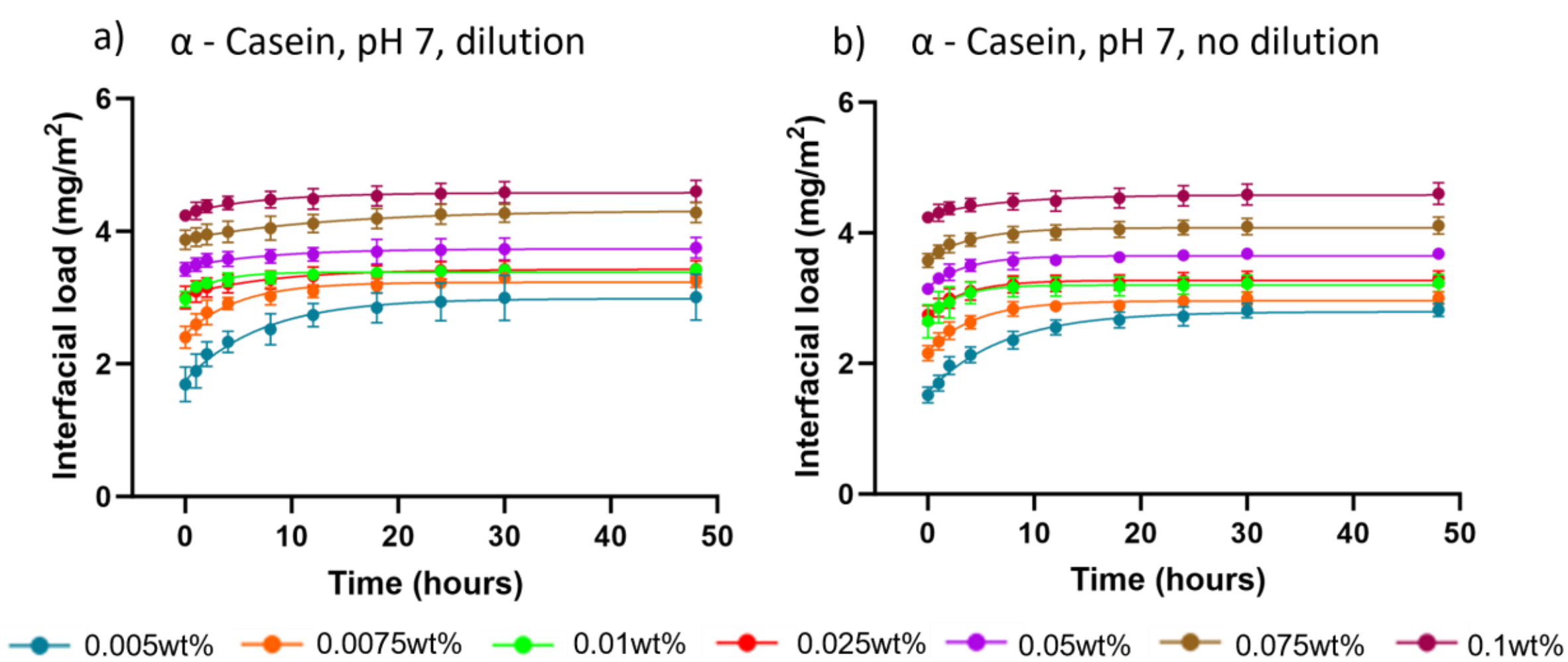

**Figure 2:** Interfacial load vs time for a) α – Casein, pH7, dilution and b) α – Casein, pH7, no dilution across 7 protein concentrations. Ionic strength is 0 mM. Solid lines represent exponential fits to the data using Equation 2 in main text. Error bars represent the standard deviation from three independent trials.

As shown in **Fig. 2**, the interfacial load of α-Casein increased over time for all concentrations tested. At lower concentrations, there was a pronounced change in the interfacial load over the 48-hour period, with significant increases occurring in the early stages of adsorption. In contrast, at higher concentrations, the initial interfacial load was already substantial, and only a slight increase in the adsorption was observed over time, suggesting that the interface was close to saturation at the initial time point. This behavior indicated a concentration dependent adsorption mechanism where α-Casein rapidly occupies available interfacial sites at lower concentrations, but slower dynamics occurred as the interface becomes saturated.

We model the adsorption kinetics using pseudo-first-order model, which incorporates parameters such as the adsorption rate constant and the equilibrium adsorption capacity.[61, 62] This model assumes that the rate of adsorption is proportional to the remaining available adsorption sites, resulting in an exponential decay in adsorption over time. Mathematically, the adsorption kinetics model is:

$$\Gamma\ (t) = \Gamma_c(1\ - e^{-Bt})\ +\ \Gamma_o \qquad (2)$$

where $\Gamma\ (t)$ represents the interfacial load ($mg/m^2$) at time t, and $\Gamma_c$, $\Gamma_o$ and $B$ are fitting parameters. Here, $\Gamma_o$ represents the interfacial load $at\ t = 0$ immediately after collection, $B$ represents the rate constant that describes how fast or slow protein adsorption occurs at the interface. As t → ∞, the saturation load, $\Gamma_\infty\ = \Gamma_o + \Gamma_c$, which suggests that $\Gamma_c$, is the differential between the saturated and initial interfacial load.

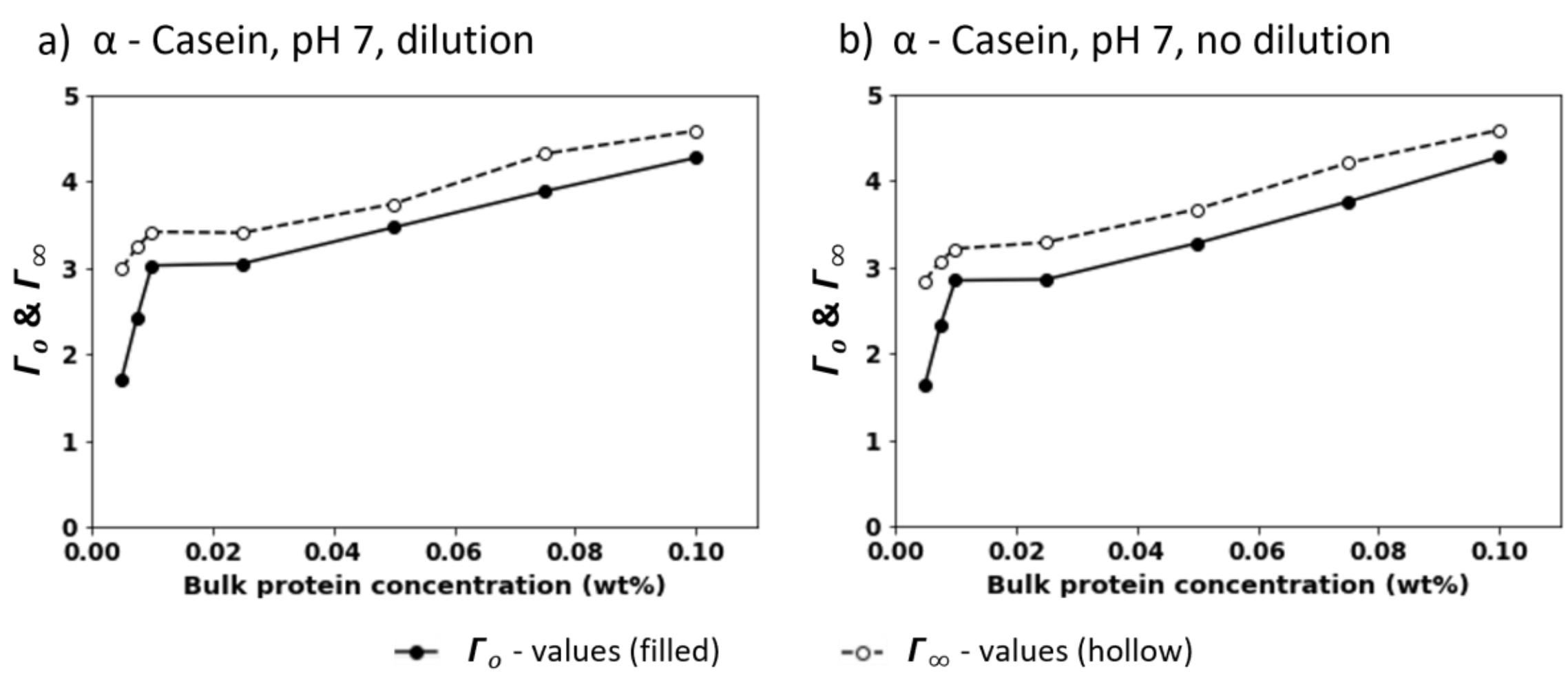

**Figure 3:** Plot of parameter values ($\Gamma_o$ & $\Gamma_\infty$) derived from equation 2 as a function of protein concentration for α – Casein with a) dilution and b) no dilution cases.

In **Fig. 2**, we fit this model to the α-Casein data sets and find that it describes well the adsorption kinetics for both the dilution and no-dilution approach. We extract the fit parameters $\Gamma_o$ and $\Gamma_\infty$ to understand how they evolve with bulk protein concentration. The initial interfacial load, $\Gamma_o$, showed an increasing trend with an increase in protein concentration **(Fig. 3a and 3b)**. Notably, a steeper increase was observed at lower concentrations, followed by a more gradual increase at higher concentrations. This trend can be attributed to the availability of free interface sites at lower concentrations, which allowed for rapid initial adsorption and interfacial coverage. In contrast, at higher concentrations, the interface was partially saturated due to the higher protein availability, leading to a smaller increase in $\Gamma_o$ as fewer unoccupied sites were available for adsorption. In contrast, $\Gamma_\infty$, representing the interfacial load after 48 hours, increased more uniformly across the concentration range **(Fig. 3a and 3b)**. Compared to $\Gamma_o$, the variation in $\Gamma_\infty$ is less pronounced at low concentrations, while higher concentrations continue to exhibit a measurable increase over time. This suggests that interfacial accumulation persists over longer durations, even after the initial adsorption stage.

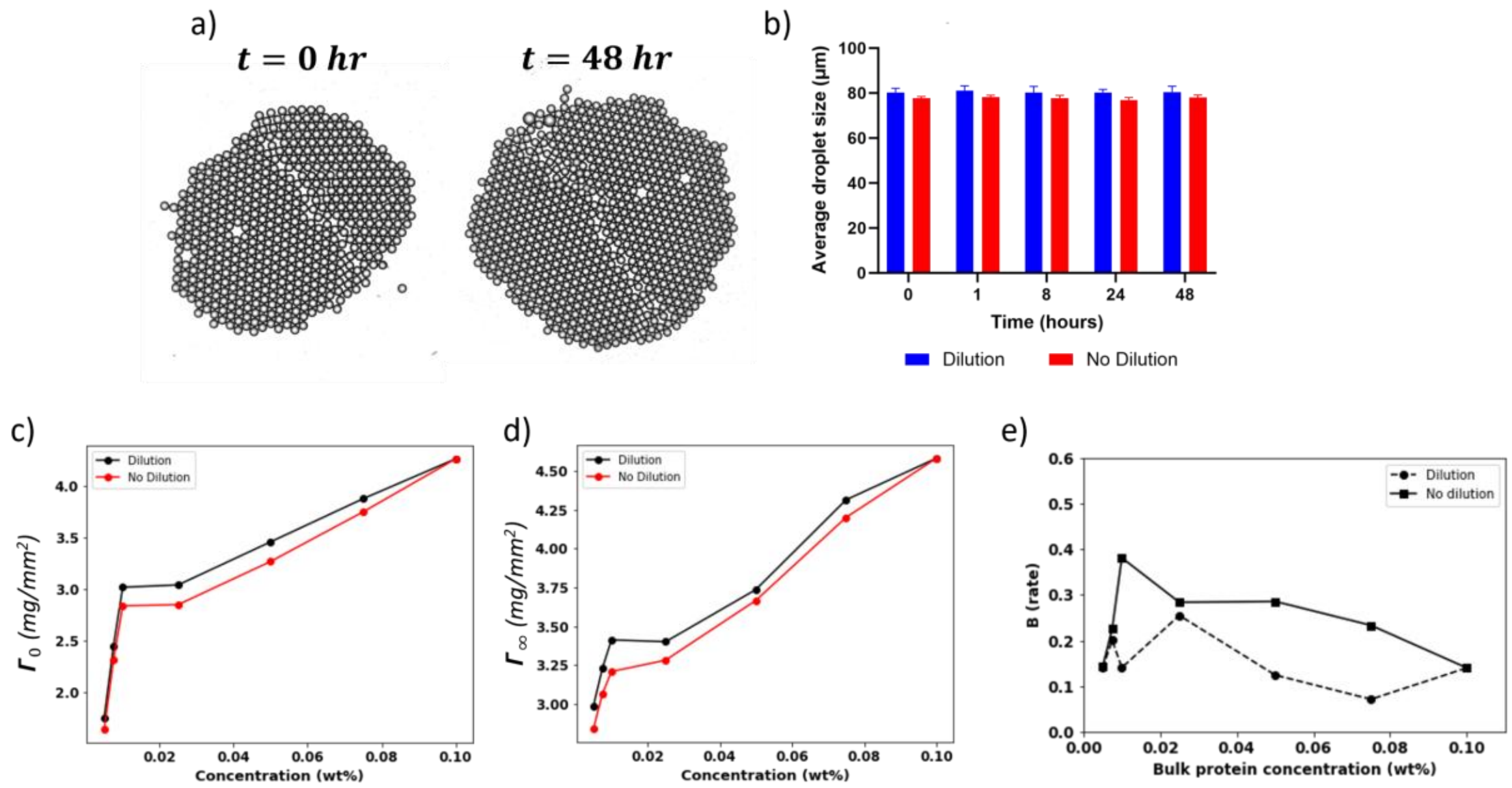

**Figure 4:** Assessment of average droplet size for α–casein-stabilized emulsions at pH7; a) Emulsions sampled at *t = 0 and t = 48* hours; b) Average droplet size w.r.t to different time points; c) Initial interfacial load $\Gamma_0$ vs protein conc for dilution and no dilution; d) ) Saturation interfacial load $\Gamma_\infty$vs protein conc for dilution and no dilution e) Parameter value *B* derived from equation 2*,* as a function of protein concentration for α – Casein with dilution and no dilution cases.

Both dilution and no-dilution techniques produced consistent droplet sizes throughout the 48-hour experiment, with diameters ranging from 75 to 80 μm, confirming the stability of the emulsification process under all conditions **(Fig. 4a,b)**. While the droplet size remained unaffected, the interfacial load showed minor differences between the two techniques **(Fig 4c,d)**. This reduction in interfacial load upon dilution is interpreted as a consequence of reduced bulk protein availability and a shift in adsorption equilibrium, rather than an explicit desorption process. The higher interfacial load observed in the dilution experiments is attributed to interfacial rearrangement and re-equilibration following dilution, rather than complete desorption of previously adsorbed proteins. The percentage difference in the initial interfacial load between the two methods ranged from $3 - 6\%$ across the concentration range **(Fig 4c)**. Likewise, the percentage difference in the saturated interfacial load between the two methods ranged from $1 - 4\%$ across the concentration range **(Fig 4d)**. These observations suggest that the dilution and no-dilution approaches to making protein-stabilized emulsions did not have a strong influence on the initial and saturated interfacial load.

The fitted $B$ values, derived from the exponential saturation model, reflect the effective progression of interfacial adsorption and interfacial rearrangement over time and showed distinct behaviors for the dilution and no-dilution experiments **(Fig. 4e)**. In the no-dilution case, the $B$ values initially increased with protein concentration, reaching a peak around 0.01 wt% before declining at higher concentrations. This trend may indicate that at lower to moderate concentrations, the proteins rapidly occupy interfacial sites, resulting in a faster approach to saturation. Beyond this point, increased crowding and potential steric hindrance could slow down the adsorption, leading to lower values at higher concentrations. In the dilution case, the values showed no consistent trend, fluctuating with no clear correlation to protein concentration. Since dilution was performed from higher concentrations, there may have been fewer available sites for protein adsorption, which could have affected the adsorption rate and contributed to the lack of a discernible trend.

In summary, for α-Casein, the interfacial adsorption kinetics is strongly dependent on bulk protein concentration and can be suitably modelled using Eqn. (2). The initial and saturated interfacial load are largely independent of which method is used to produce emulsions; however, the adsorption rate is influenced by the choice of making emulsions. In the remaining sections, we use the dilution method for making emulsions to avoid wettability issues and focus on characterizing the interfacial adsorption kinetics using the initial interfacial load $\Gamma_o$ and the saturation load, $\Gamma_\infty$.

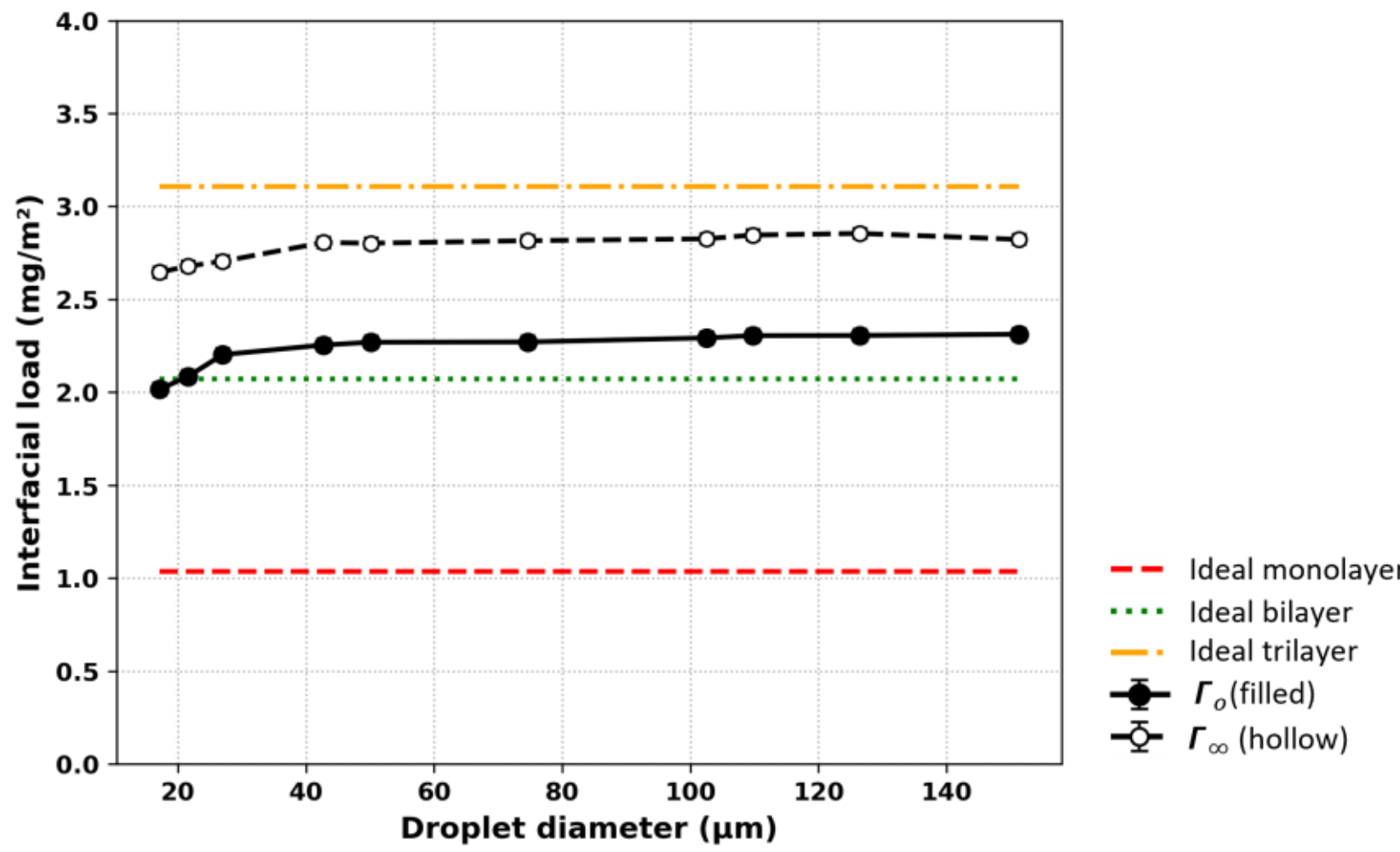

**Figure 5:** Interfacial load $\boldsymbol{\Gamma}_0$ (filled circles) and $\Gamma_\infty$ (hollow circles) as a function of droplet diameter. Colored reference lines indicate idealized surface coverage based on geometric packing assumptions: monolayer (red, dashed), bilayer (green, dotted), and trilayer (orange, dash–dot). Error bars represent standard deviation and are smaller than the marker size.

## ii. Effect of droplet size on interfacial protein load

Protein adsorption measurements in emulsions are inherently sensitive to droplet size because interfacial load is determined from the adsorbed protein mass normalized by the total available oil–water interfacial area. In conventional emulsification processes, droplet populations are typically highly polydisperse, with reported droplet diameters spanning from submicron scales to tens of microns depending on formulation and processing conditions[4, 35]. Consequently, the total interfacial area depends not only on the mean droplet size but also on the underlying droplet size distribution, complicating accurate determination of the surface area available for adsorption. Since depletion-based adsorption measurements rely directly on the total interfacial area, variability in droplet size distributions can propagate into the calculated interfacial load and potentially obscure the influence of variables such as protein type, pH, and ionic strength. Beyond its geometric influence on total interfacial area, droplet size also determines interfacial curvature, which has been reported to affect protein adsorption and molecular organization at curved interfaces[63-66]. Monodisperse emulsions provide a controlled means of examining this dependence because the total interfacial area can be directly related to the measured droplet diameter. To investigate the influence of droplet size on interfacial loading, we generated monodisperse α-casein emulsions with droplet diameters spanning approximately $16 - 150\ \mu m$ at two protein concentrations (0.01 and 0.1 wt%).

Based on simple geometric considerations, interfacial load would be expected to remain independent of droplet size if adsorption were governed solely by the available interfacial area. For a fixed oil volume, subdivision into progressively smaller droplets increases the total interfacial area and, assuming sufficient protein availability, the total adsorbed protein mass proportionally, such that the calculated interfacial load unchanged (**Table S2**). At 0.01 wt% α-Casein, the initial post-collection interfacial load $\Gamma_o$ remained approximately constant for droplets larger than $\sim 40\ \mu m$, whereas a modest reduction was observed for smaller droplets ($17.4,\ 21.6,$ and $27\ \mu m$) (**Fig 5**). In contrast $\Gamma_\infty$, showed no statistically resolved dependence on droplet diameter across the investigated range. The decrease in $\Gamma_o$ between the largest and smallest droplets was approximately $12.8\%$, compared with approximately $6.3\%$ for $\Gamma_\infty$. Thus, the observed droplet-size-associated variation was more pronounced in the early interfacial load than in the long-time loading.

To further examine this behavior, the droplet-size experiments were repeated at a tenfold higher α-casein concentration (0.1 wt%). Under these conditions, both $\Gamma_o$ and $\Gamma_\infty$ remained approximately constant across the investigated droplet-size range (**Fig. S3**). This additional experiment indicates that the variation observed at 0.01 wt% depends on protein concentration and is not a consistent feature of the measured interfacial load across both concentrations. Because smaller droplets generate a greater total interfacial area for a fixed dispersed-phase volume, protein availability and depletion from the aqueous phase were also considered as possible contributing factors. However, the measured residual serum protein concentrations ($C_{ser}$) did not reveal a systematic relationship sufficient to establish bulk depletion as the cause of the observed variation (**Table S3**). Furthermore, the investigated droplet-size range was obtained using different microfluidic device geometries and flow conditions, which may also influence the early adsorption history (See **Fig S1**). Consequently, the modest variation in $\Gamma_o$ cannot be attributed uniquely to interfacial curvature or protein depletion. Although curvature-dependent adsorption and molecular organization have been reported in the literature, the molecular-scale curvature effects examined in those studies cannot be directly extrapolated to the micrometer-sized droplets investigated here.

The magnitude of the measured interfacial loads was further compared with idealized geometric estimates of protein coverage. Using a simplified disc-packing representation, in which each adsorbed protein occupies an interfacial area of $\pi R^2$, where $R$ is the hydrodynamic radius of the protein. For α-Casein ($R \approx 3\ nm$)[56], this simple geometric estimate yields an ideal monolayer coverage of approximately 1 $mg/m^2$, substantially lower than the experimentally measured $\Gamma_o$ and $\Gamma_\infty$ values. Even under more permissive packing assumptions, the measured interfacial loads remain above those predicted for a simple close-packed monolayer. We also evaluated idealized bilayer and trilayer packing limits, defined as two and three times the estimated monolayer coverage, respectively, and found that the experimentally measured loads lie between these two bounds across the investigated droplet-size range **(Fig. 5).** Although these geometric estimates do not establish the formation of discrete multilayers, they indicate protein accumulation beyond that predicted by a simplified monolayer model, potentially reflecting conformational rearrangement, alternative packing arrangements, or additional protein accumulation at the interface.

These findings have practical implications for quantifying protein adsorption in real emulsions, where polydispersity introduces uncertainty in the determination of total interfacial area[1, 18, 20-22]. In contrast, the monodisperse emulsions generated here enable well-defined interfacial-area calculations and systematic evaluation of droplet-size-associated variations in protein adsorption. Although a modest reduction in interfacial load was observed for the smallest droplets at 0.01 wt% α-Casein, this variation was less pronounced at longer adsorption times and was not observed at 0.1 wt%. Overall, these findings highlight the importance of accounting for droplet size and protein concentration when interpreting interfacial loading and demonstrate the utility of monodisperse microfluidic emulsions for quantitative adsorption measurements.

### iii. Scaling of protein concentration dependence on interfacial adsorption

Bulk protein concentration is expected to play a central role in interfacial adsorption because it governs the amount of protein available for interfacial accumulation[14]. To examine whether the measured interfacial load could be described through a simple concentration-dependent relationship, α-Casein was used here as a representative model protein. The measured $\Gamma_o$ values were analyzed as a function of bulk concentration (0.05 – 0.1 wt) across different pH (= 3, 5, 7) and ionic strength (= 0, 0.1 M) conditions. We found minor effect of pH and ionic strength on $\Gamma_o$ values and pooled the data from different conditions to assess the scaling with α-Casein concentration.

α-Casein data follow a clear concentration-dependent trend despite spanning multiple solution conditions (**Fig 6**). This indicates that bulk concentration acts as the dominant variable governing the measured interfacial load, while variations in pH and ionic strength introduce secondary deviations around this broader relationship. The resulting fit yields a scaling exponent of approximately $0.21 \pm 0.01$, indicating that the initial interfacial load increases in a sublinear manner across the tested concentration range.

Such sublinear scaling indicates a transition from bulk-driven adsorption to interface-limited behavior. At higher concentrations, the interface becomes progressively occupied, and further adsorption is increasingly constrained by crowding, steric interactions, and non-ideal packing within the interfacial layer [2, 15]. In this regime, additional protein arriving from the bulk contributes less effectively to increases in surface excess, reflecting the growing influence of lateral interactions and limited interfacial accommodation. This explanation is also consistent with our finding that $\Gamma_\infty$ is largely independent of protein bulk concentration ($\Gamma_\infty \sim C^{0.12}$, see **Fig. S4**).

Overall, these results demonstrate that the initial interfacial load $\Gamma_o$ for α-casein can be effectively organized through a simple concentration-dependent scaling, with bulk concentration acting as the primary controlling variable and solution chemistry introducing secondary modulation.

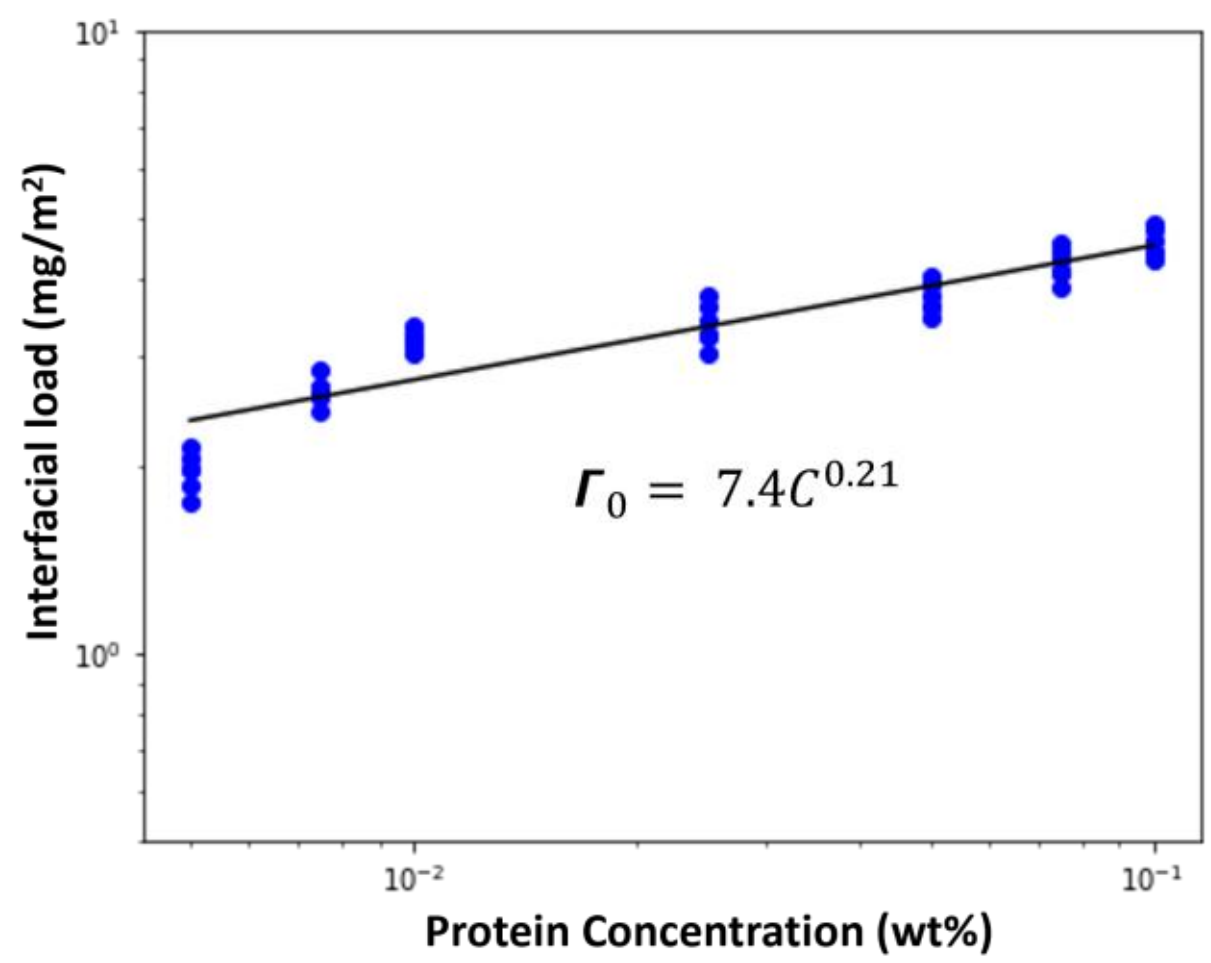

**Figure 6:** Scaling of initial interfacial load ($\Gamma_0$) with protein concentration for α-casein, pooled across all pH and ionic strength conditions. Solid line represents a power-law fit to the data, with the corresponding fitted equation and parameters shown in the plot.

### iv. High-throughput mapping of protein adsorption across environmental conditions

Interfacial protein adsorption depends on multiple factors, including bulk protein concentration, pH, ionic strength, and protein identity. Mapping such large experimental parametric space using conventional methods of homogenized emulsions are not only laborious but also quite limiting when available protein samples are small. Here, we leverage our microfluidic workflow, and systemically map adsorption behavior across a multidimensional parameter space spanning three proteins (α-Casein, β-Casein, and WPI), seven concentrations, three pH values, and two ionic strength conditions (42 conditions per protein). The resulting high-throughput dataset is visualized as heatmaps of interfacial load at $\Gamma_0$ and $\Gamma_\infty$ (**Fig. 7**), enabling direct comparison of adsorption behavior across all tested conditions.

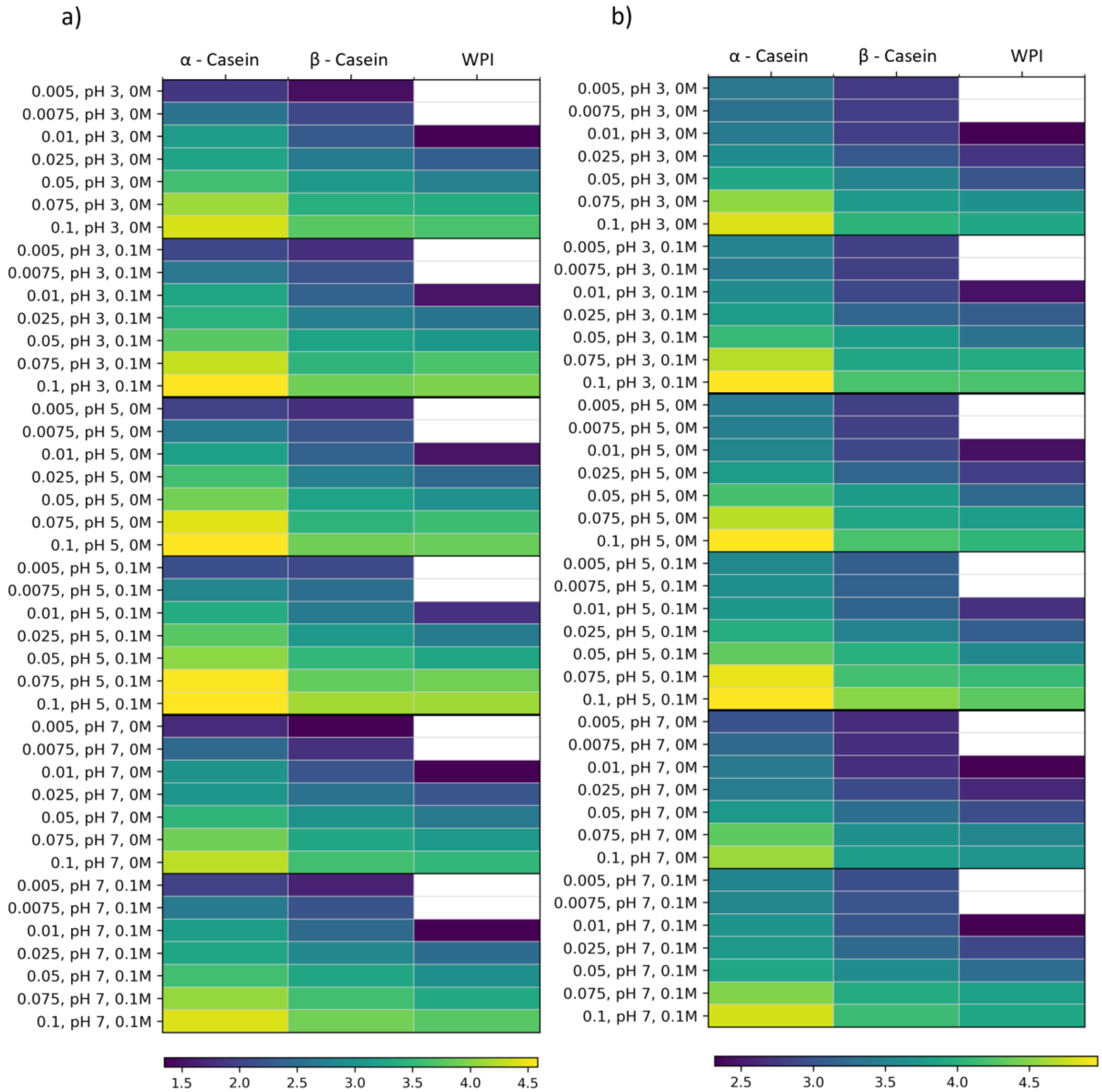

**Figure 7:** Heatmap representation of interfacial load at (a) $\Gamma_0$ and (b) $\Gamma_\infty$ for three proteins (α-casein, β-casein, and WPI). Columns correspond to protein type, and rows represent 42 conditions per protein (seven concentrations, three pH values, and two ionic strengths). The color scale denotes the magnitude of interfacial load, highlighting trends across concentration, pH, and ionic strength.

As shown in **Fig. 7**, a clear global trend of increasing interfacial load with increasing bulk protein concentration was observed across all proteins and solution conditions. This trend was consistent at both $\Gamma_o$ and $\Gamma_\infty$ , reinforcing the dominant role of protein concentration in governing interfacial adsorption compared to pH and ionic strength. Superimposed on this concentration-driven behavior, systematic differences between protein types are evident, with α-Casein consistently exhibiting higher interfacial loads than β-Casein and WPI under comparable conditions. For example, at pH 7, α-Casein exhibited interfacial loads approximately $\sim 15 - 25\%$ higher than β-Casein and $\sim 18 - 60\%$ higher than WPI across the tested concentration range at $\Gamma_o$ , with the magnitude of the difference decreasing with increasing concentration. Notably, the difference between β-casein and WPI is comparatively smaller across conditions ($\sim 5 - 30\%$), with the smallest differences observed at higher protein concentrations, suggesting that interfacial coverage becomes increasingly limited by protein availability and packing constraints rather than molecular structure. These differences reflect the underlying structural characteristics of the proteins. Flexible proteins such as α-Casein can readily adapt to the interface and achieve higher surface coverage, whereas more compact, globular proteins such as WPI exhibit more limited adsorption due to the need for conformational rearrangement prior to interfacial accommodation[12, 13]. As a result, while concentration governs the overall magnitude of interfacial load, protein structure defines the upper limit of achievable adsorption under a given set of conditions.

The variations in pH introduce a systematic secondary modulation to the protein adsorption behavior. Higher interfacial loads were consistently observed at pH 5 (the isoelectric point of the proteins), while conditions further from the pI exhibit comparatively lower loads. For example, at 0.025wt% α - Casein, the $\Gamma_o$ at pH 5 was approximately $\sim 10\%$ and $15\%$ higher than that measured at pH 3 and pH 7 respectively. This trend was observed across all three proteins and persists at both $\Gamma_o$ and $\Gamma_\infty$. Near the isoelectric point, proteins carry minimal net charge, reducing electrostatic repulsion and enabling closer packing at the interface. In contrast, at pH values far from the pI, increased charge leads to stronger repulsive interactions, limiting the extent of adsorption and resulting in lower interfacial loads.

In addition to pH, ionic strength exerted a clear and consistent influence on interfacial adsorption across all proteins and conditions. Increasing ionic strength led to higher interfacial loads at comparable concentrations and pH values, with increases ranging from ~5% to ~15% between 0 and 0.1 M salt across all proteins and conditions. This trend was evident across both $\Gamma_o$ and $\Gamma_\infty$ and was consistent for α-Casein, β-Casein, and WPI. The observed increase in adsorption with ionic strength was attributed to electrostatic screening effects, where the presence of salt reduces repulsive interactions between charged protein molecules at the interface, enabling denser packing and greater surface coverage. As a result, higher ionic strength facilitated increased interfacial load by promoting more efficient accommodation of proteins within the adsorbed layer. Additional measurements on β-lactoglobulin (BLG) are provided in the Supporting Information (**Fig. S5**) and are not included in the main analysis to maintain focus on the primary protein systems studied.

## 4. Conclusions

This study presents a comprehensive investigation of protein adsorption at the oil–water interface using a high-throughput microfluidic platform capable of quantifying interfacial protein load across a wide range of concentrations, pH values, and ionic strengths. By combining controlled droplet generation with time-resolved adsorption measurements, we extracted early ($\Gamma_o$) and saturation ($\Gamma_\infty$) interfacial loads for structurally distinct proteins under well-defined formulation conditions.

Interfacial adsorption was quantified following droplet generation and collection, with time-dependent evolution of interfacial load described using a pseudo-first-order model. The measured $\Gamma_o$ and $\Gamma_\infty$ were similar for dilution and no-dilution approaches (within ~ $1 - 6\%$), indicating minimal influence of emulsification method on the measured interfacial load. In contrast, the adsorption rate parameter ($B$) showed greater variability, suggesting sensitivity to interfacial crowding and rearrangement.

A key uncertainty in interfacial load measurements in emulsions arises from the use of average droplet sizes and the assumption that adsorption scales solely with total interfacial area, despite the inherently polydisperse nature of conventional emulsification processes. To examine droplet-size-associated variations under well-defined interfacial-area conditions, monodisperse emulsions spanning approximately 17–151 μm were investigated. At 0.01 wt% α-casein, a modest reduction in $\Gamma_o$ was observed for the smallest droplets, whereas $\Gamma_\infty$ showed no statistically significant dependence on droplet diameter. At 0.1 wt%, both $\Gamma_o$ and $\Gamma_\infty$ remained approximately independent of droplet size. These observations indicate that the measured interfacial load is comparatively insensitive to droplet diameter across the investigated range, although modest early-loading variations may arise under dilute protein conditions. Comparison with idealized interfacial packing limits further revealed that the measured loads exceeded simplified monolayer estimates and fell between bilayer and trilayer reference values, suggesting protein accumulation beyond that predicted by a simple monolayer packing model without establishing a specific multilayer structure.

Scaling analysis further demonstrates that bulk protein concentration serves as the primary organizing variable governing interfacial load, with both $\Gamma_o$ and $\Gamma_\infty$ exhibiting sublinear scaling behavior. This sublinear response indicates diminishing increases in interfacial load at higher protein concentrations, potentially reflecting crowding, steric constraints, and non-ideal packing increasingly restrict additional adsorption at higher concentrations. High-throughput mapping across proteins and solution conditions reveals that protein structure, pH, and ionic strength introduce systematic but secondary modulation to adsorption behavior. Flexible proteins such as α-casein consistently achieve higher interfacial loads than β-casein and WPI, while adsorption is enhanced near the isoelectric point and at elevated ionic strength due to reduced electrostatic repulsion and improved packing efficiency.

Overall, this work establishes a quantitative framework for measuring protein adsorption under controlled interfacial conditions. The combination of precise droplet generation and high-throughput depletion analysis enables reliable comparison of protein functionality while reducing uncertainties associated with droplet polydispersity and interfacial area estimation. This approach provides a practical route for systematic screening of protein-based emulsifiers and for linking formulation conditions to interfacial adsorption behavior.

## Conflict of interest

There are no conflicts of interest to declare.

## Data availability

The data supporting this article have been included as part of the Supplementary Information.

## Acknowledgements

We gratefully acknowledge Motif Foodworks Inc for their financial support for this research (Grant no. 20-0159).

**High-throughput microfluidic quantification of kinetics of interfacial protein load in monodisperse oil-in-water emulsions**

Varun Potdar[1], Alexander Favela[1], Michael Boehm[2], Stefan Baier[3], Siva Vanapalli[1]

# Supplementary Figures

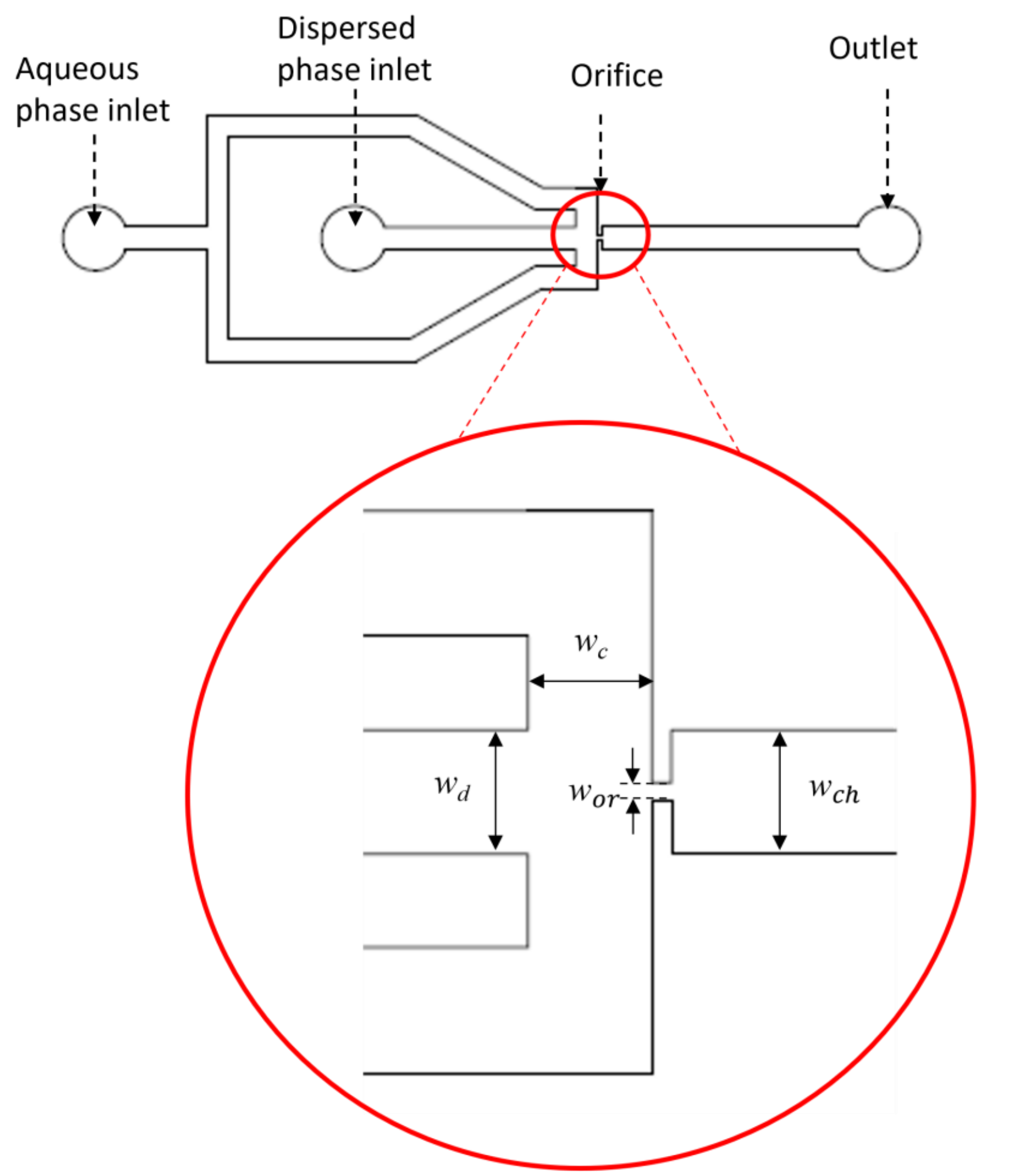

| Geometrical variables | | $w_{or}$ | $w_{ch}$ | $w_d$ | $w_c$ | $h$ |
|---|---|---|---|---|---|---|
| Channel Dimensions (µm) | FFD-1 | 50 ± 3 | 296 ± 5 | 298 ± 4 | 297 ± 4 | 100 ± 4 |
| | FFD-2 | 45 ± 2 | 70 ± 2 | 65 ± 2 | 64 ± 2 | 100 ± 2 |
| | FFD-3 | 25 ± 2 | 240 ± 3 | 80 ± 3 | 80 ± 3 | 50 ± 2 |

**S1:** Schematic representation of the microfluidic flow-focusing device used for monodisperse droplet generation.

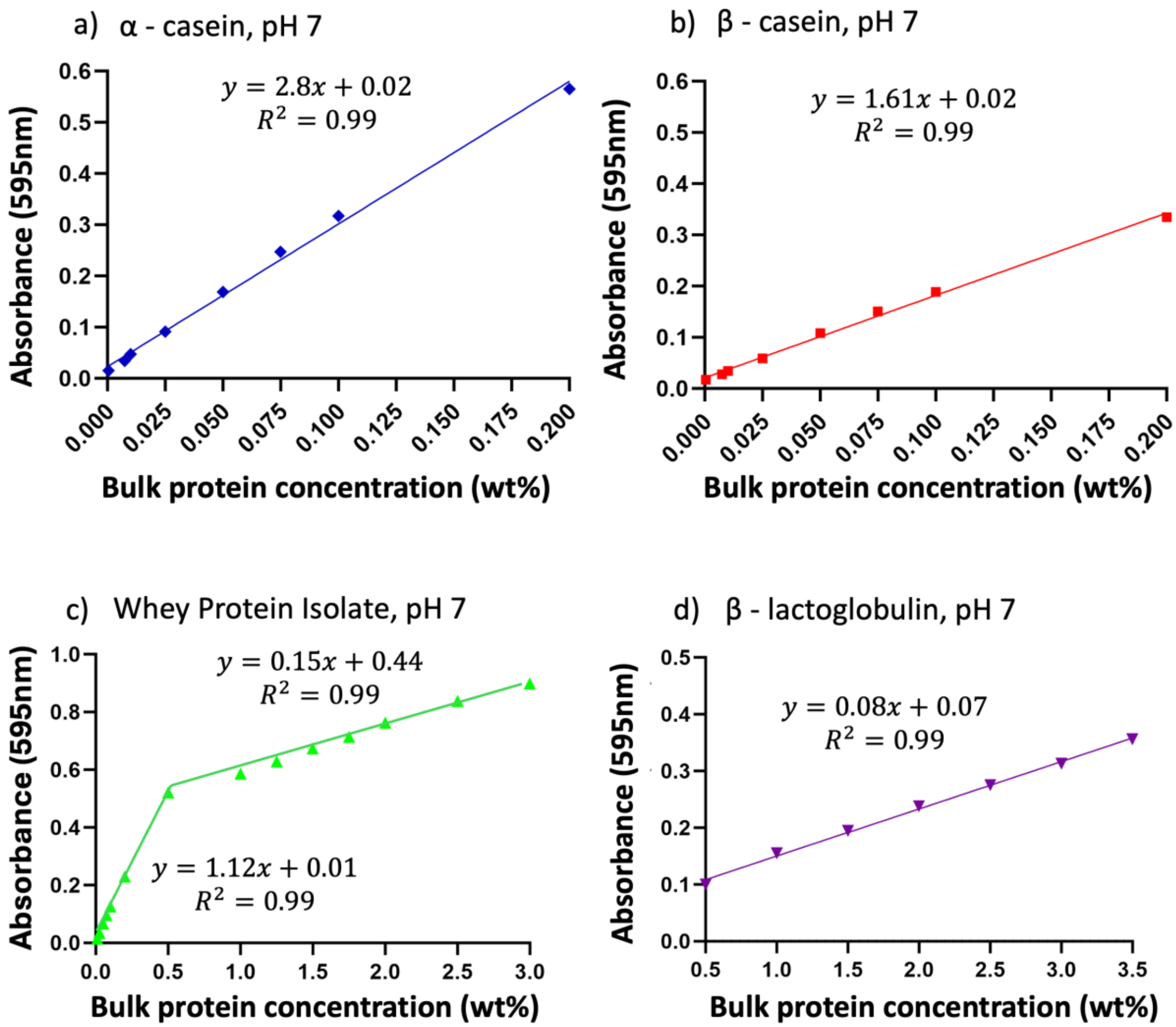

**S2:** Absorbance values vs Protein concentration for a) α-casein; b) β-Casein; c) WPI low concentrations; d) WPI high concentration; e) B-lactoglobulin. Each plot indicates a linear fit with the corresponding equation and $R^2$ value.

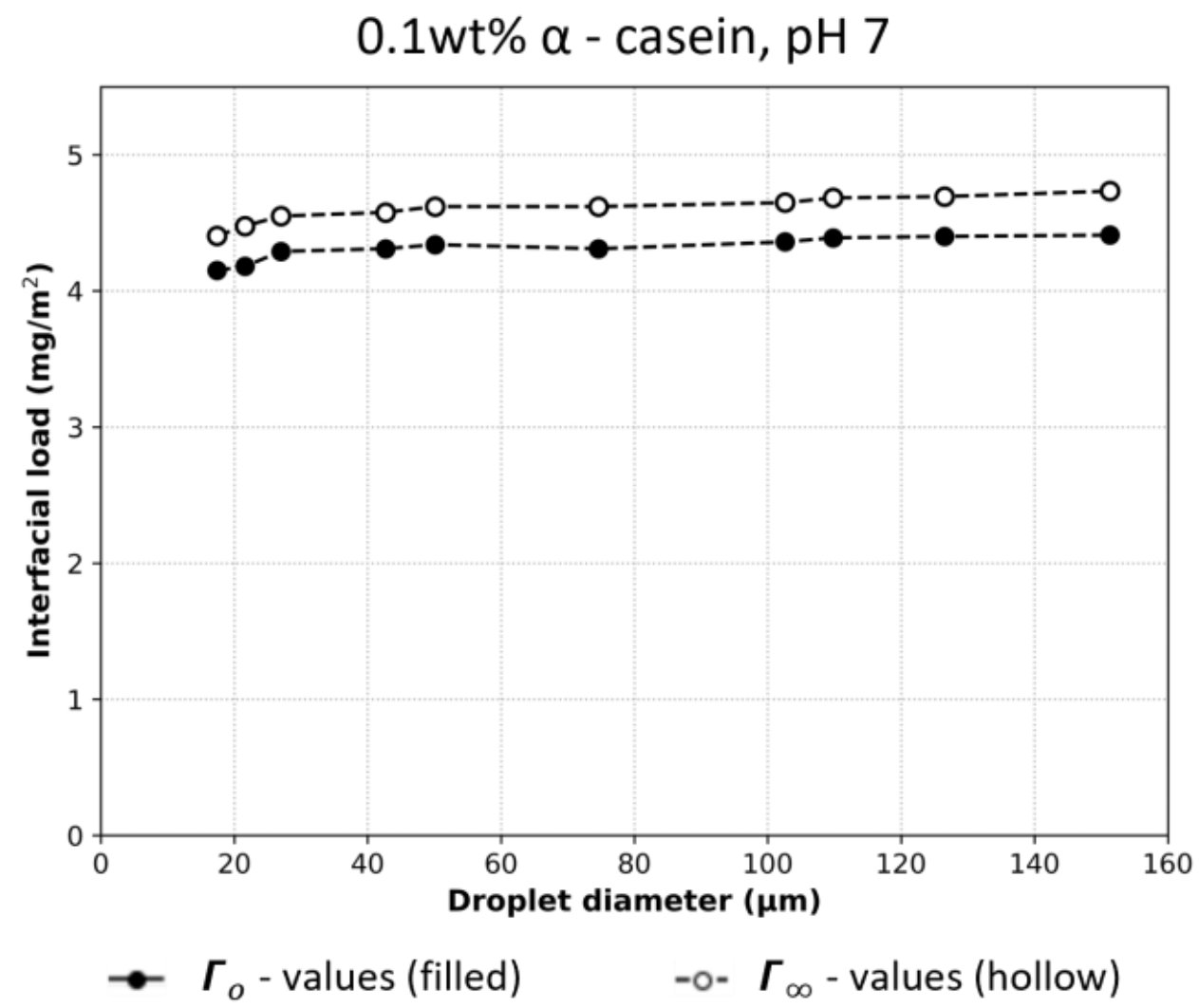

**S3:** Interfacial load $\boldsymbol{\Gamma}_0$ (filled circles) and $\Gamma_\infty$ (hollow circles) as a function of droplet diameter.

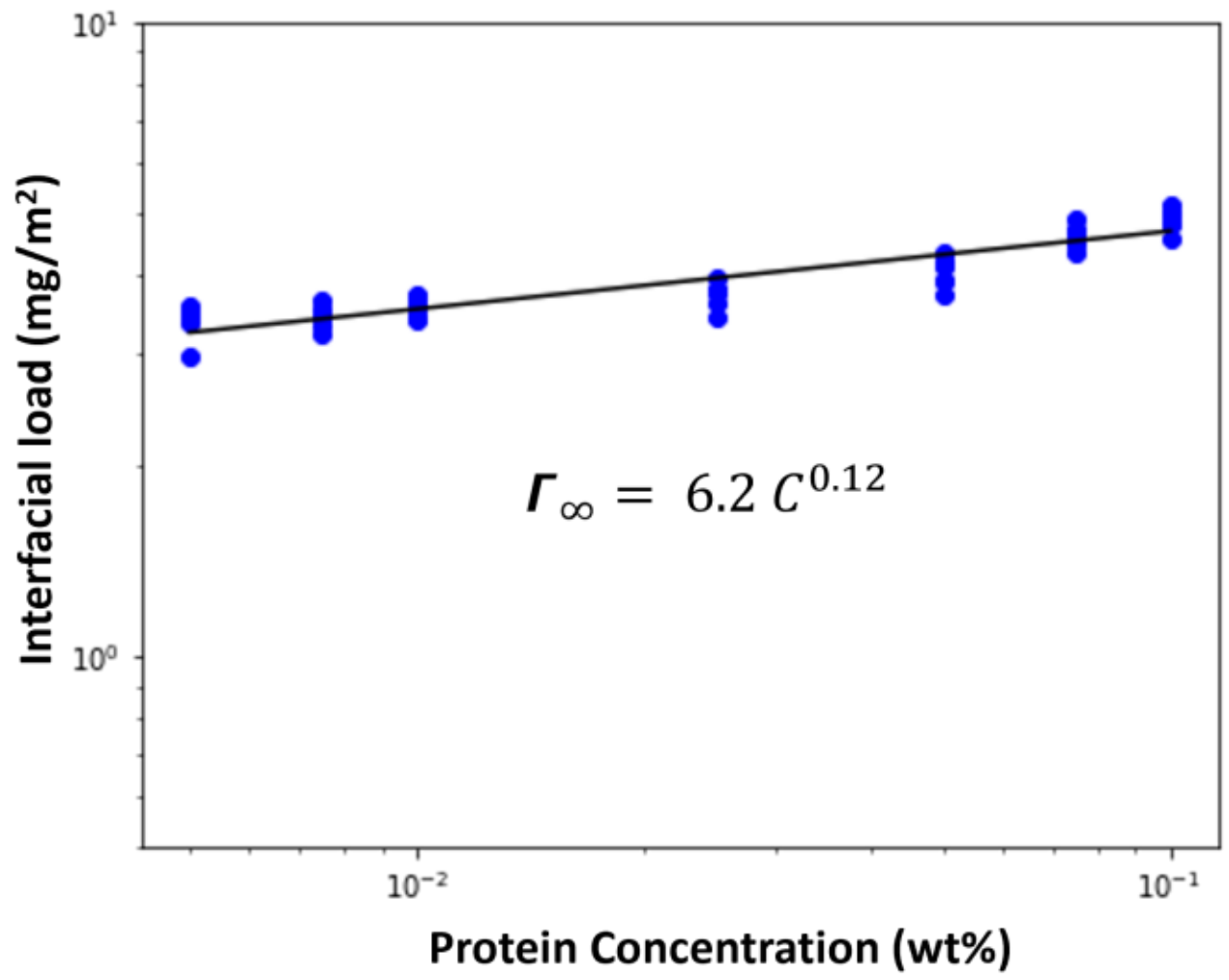

**S4:** Scaling of saturation load ($\Gamma_\infty$) with protein concentration for α-casein, pooled across all pH and ionic strength conditions. Solid line represents a power-law fit to the data, with the corresponding fitted equation and parameters shown in the plot.

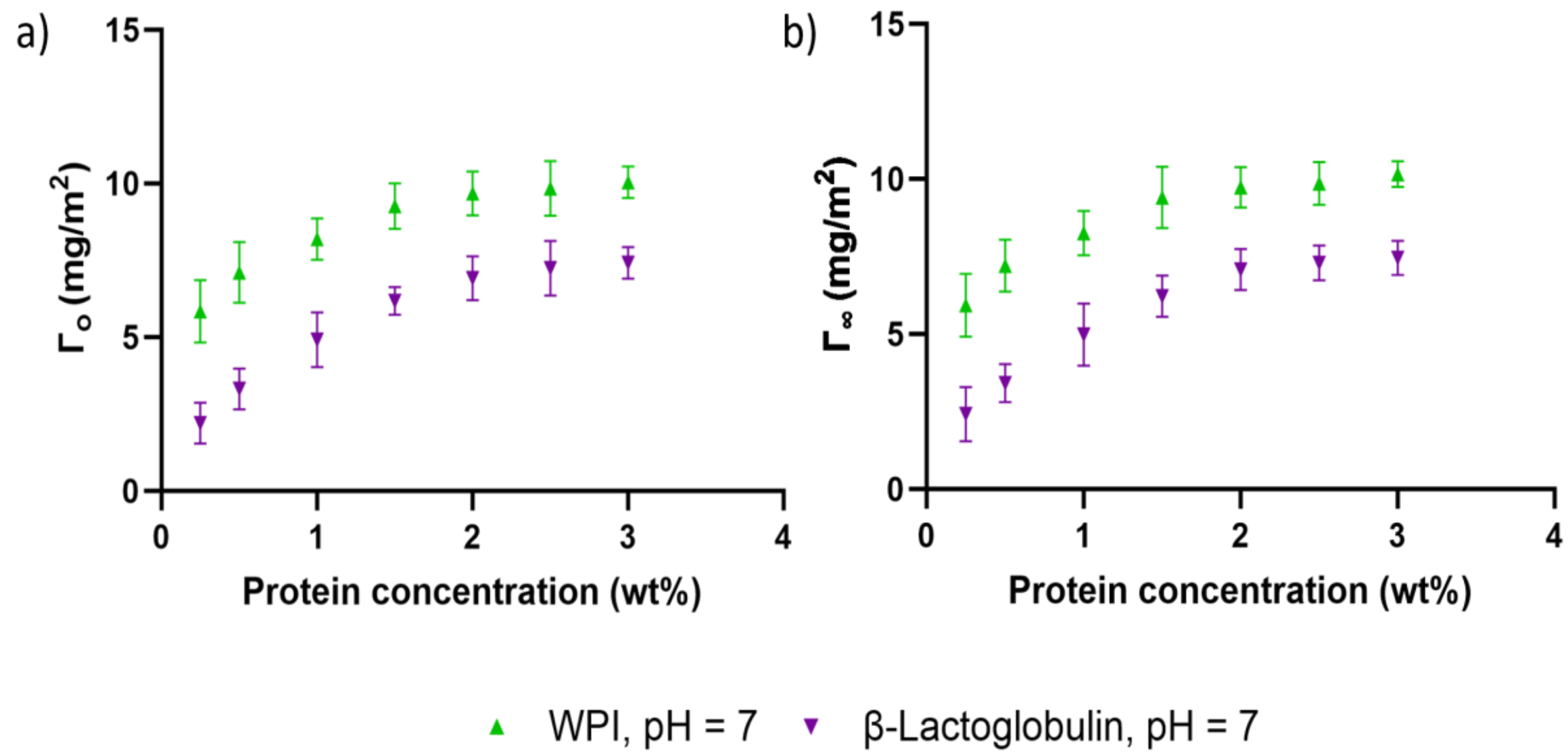

**S5:** Interfacial load vs protein concentration for WPI and β – Lactoglobulin; a) $\Gamma_0$ at pH7; b) $\Gamma_\infty$ at pH7

| Proteins | Protein Conc range (wt%) | Oil phase | pH | Emulsification Technique | Droplet size ($um$) | $\Gamma\left(\frac{mg}{m^2}\right)$ | Study |
|---|---|---|---|---|---|---|---|
| Mixture of α-Casein and β-Casein | 0.5 | n-tetradecane | 7 | Valve homogenizer | 1-2 | 3 | Dickinson et al.[1] |
| β-Lactoglobulin/Tween 20 | 0.5 | n-tetradecane | 7 | Valve homogenizer | 1-2 | 1.7 | Courthaudon et al.[2] |
| Whey Protein Isolate | 0.3-2.5 | Soya | 7 | Microfluidizer | 1-5 | 1.5 - 3.2 | Hunt et al.[3] |
| Sodium Caseinate | | | 7 | | | 1 - 3.5 | |
| β-Casein | 0.1 - 1.5 | Soya | 6.7 | Valve homogenizer | 1-3 | 1 - 4 | Smuders[4] |
| β-Lactoglobulin | 0.1 - 2.5 | | 6.7 | | 1-2 | 1 - 2 | |
| Ovalbumin | 0.1 - 2.5 | | 6.7 | | 1-4 | 1 - 2 | |
| β-Lactoglobulin | 0.01 - 0.1 | Soybean | 6.2 | Rotor-stator homogenizer | 15-25 | 1.5 - 3 | Tcholakova et al.[5] |
| β-Casein | 0.5 | MCT oil | 7 | Mixer | 2 | 2.6 - 2.7 | Waninge et al.[6] |
| β-Lactoglobulin | 0.5 | | | | 1.5 | 2.3 - 2.7 | |
| Whey Protein Isolate | 0.2 - 2 | Sunflower oil | 7 | Microfluidizer | 0.2 - 2 | 2 - 6 | Hinderink et al.[7] |
| Sodium Caseinate | 0.2 - 2 | Sunflower oil | 7 | | 0.2 - 1 | 1 - 4 | |

**Table S1:** Literature studies with the experimental conditions, emulsification techniques and interfacial load values for proteins.[1-7]

**Illustrative geometric scaling of interfacial area with droplet size:**

To further illustrate the rationale underlying the droplet-size analysis presented in Section 3.2 of the main manuscript, an idealized geometric scaling analysis is provided in Table S1. Assuming spherical, monodisperse droplets and a fixed interfacial protein load ($\Gamma$), decreasing the droplet diameter increases the total interfacial area for a given dispersed-phase volume. Consequently, the total amount of adsorbed protein increases proportionally with the available interfacial area, while the interfacial load (adsorbed protein per unit area) remains unchanged. This simplified analysis is intended only to demonstrate the geometric relationship between droplet size, interfacial area, and adsorbed protein mass, and does not represent experimental measurements.

| Diameter (um) | Relative number of drops | Total interfacial area ($m^2$) | Total adsorbed protein mass (mg) | Interfacial load ($mg/m^2$) |
|---|---|---|---|---|
| 100 | 1 | 1 | 1 | 1 |
| 50 | 8 | 2 | 2 | 1 |
| 25 | 64 | 4 | 4 | 1 |
| 10 | 1000 | 10 | 10 | 1 |
| 1 | 1000000 | 100 | 100 | 1 |

**Table S2:** Illustrative geometric scaling showing the effect of droplet diameter on the relative number of droplets, total interfacial area, total adsorbed protein mass, and normalized interfacial load for monodisperse emulsions. Values are normalized relative quantities and are presented solely to demonstrate the geometric scaling relationship assuming a fixed interfacial protein load.

**Assessment of protein depletion across droplet sizes**

To assess whether differences in bulk protein availability could contribute to the droplet-size-associated variation in initial interfacial load ($\Gamma_o$), residual protein concentration in the continuous phase after emulsification ($C_{ser}$) were examined for α-casein emulsions prepared at an initial concentration ($C_{ini}$) of 0.01 wt%. The percentage protein depletion was calculated as:

$$Depletion\ (\%) = \frac{(C_{ini} - C_{ser})}{C_{ini}} \times 100$$

The corresponding values are summarized in Table S3. Protein depletion varied across the investigated droplet diameters but did not exhibit a systematic relationship with droplet size or the observed reduction in $\Gamma_o$ for the smallest droplets. Although these measurements quantify the reduction in aqueous protein concentration, they do not establish bulk depletion as the primary cause of the observed droplet-size-associated variation in initial interfacial loading.

| Droplet dia (µm) | $\Gamma_o$ (mg/m²) | $C_{ini}$ (wt%) | $C_{ser}$ (wt%) | Protein depletion (%) |
|---|---|---|---|---|
| 17.4 | 2.0156 | 0.01 | 0.009969 | 0.31 |
| 21.6 | 2.0855 | 0.01 | 0.00972 | 2.8 |
| 27 | 2.2025 | 0.01 | 0.009465 | 5.35 |
| 42.7 | 2.254 | 0.01 | 0.009904 | 0.96 |
| 50.1 | 2.2692 | 0.01 | 0.009731 | 2.69 |
| 74.6 | 2.2705 | 0.01 | 0.008739 | 12.61 |
| 102.6 | 2.2925 | 0.01 | 0.007461 | 25.39 |
| 109.8 | 2.3046 | 0.01 | 0.007076 | 29.24 |
| 126.5 | 2.3055 | 0.01 | 0.009698 | 3.02 |
| 151.3 | 2.3122 | 0.01 | 0.009259 | 7.41 |

**Table S3:** Initial post-collection interfacial protein load ($\Gamma_o$), residual serum protein concentration (Cser), and percentage protein depletion for monodisperse α-casein emulsions (0.01 wt%) generated at different droplet diameters.